\documentstyle[12pt]{article}
\newcommand{\be}{\begin{equation}}
\newcommand{\bea}{\begin{eqnarray}}
\newcommand{\eea}{\end{eqnarray}}
\newcommand{\ba}{\begin{array}}
\newcommand{\ea}{\end{array}}
\newcommand{\ee}{\end{equation}}

\expandafter\ifx\csname mathbbm\endcsname\relax

\else

\fi \textheight 22cm \textwidth 15cm \topmargin 1mm \oddsidemargin
5mm \evensidemargin 5mm

\begin{document}
\begin{titlepage}
\hfill
\vbox{
    \halign{#\hfil         \cr
           hep-th/0207037 \cr
           IPM/P-2002/028 \cr
           } 
      }  
\vspace*{20mm}
\begin{center}
{\Large {\bf On Type IIA String Theory on the PP-wave Background}\\ }

\vspace*{15mm} \vspace*{1mm} {Mohsen Alishahiha$^a$, Mohammad A.
Ganjali$^b$, Ahmad Ghodsi$^{a,b}$\\ and Shahrokh Parvizi$^a$}

\vspace*{1cm}

{\it $^a$ Institute for Studies in Theoretical Physics and Mathematics (IPM)\\
P.O. Box 19395-5531, Tehran, Iran\\
\vspace*{1mm}
$^b$ Department of Physics, Sharif University of Technology\\
P.O. Box 11365-9161, Tehran, Iran}\\

\vspace*{1cm}
\end{center}

\begin{abstract}
We study type IIA superstring theory on a PP-wave background with 24
supercharges. This model can exactly be solved and then quantized.
The open string in this PP-wave background is also studied. We
observe that the theory has supersymmetric Dp-branes for
$p=2,4,6,8$.
\end{abstract}

\end{titlepage}

\section{Introduction}
An interesting observation in \cite{MET} (see also \cite{MT})
is that type
IIB string theory on the 10-dimensional PP-wave background
with five form RR field strength is exactly solvable. In fact
it has been shown that in the light-cone gauge the
Green-Schwarz action of type IIB string theory
in this background is quadratic in both bosonic and fermionic
sectors and can therefore be quantized in spite of the
presence of RR field. Besides it has also been shown
\cite{BFHP} that this background is a maximally
supersymmetric solution of type IIB supergravity.
Thus with the flat ten dimensional Minkowski space and
$AdS_5\times S^5$ we have three maximally supersymmetric
background of type IIB string theory but two of them have
led to the exactly solvable models so far.

Another interesting fact is that this maximally
supersymmetric PP-wave background can be obtained as a Penrose
limit of another maximally supersymmetric background which
is $AdS_5\times S^5$ \cite{Blau2}. On the other hand type IIB
string theory on $AdS_5\times S^5$ is believed to be dual to
the ${\cal N}=4$ SYM theory in four dimensions. Indeed it has
been shown \cite{BMN} that taking Penrose limit of
$AdS_5\times S^5$ has a corresponding limit for the gauge theory
as well. By making use of this fact the authors of \cite{BMN}
have been able to identify the excited string states with a
class of gauge invariant operators in the large $N$ limit of
${\cal N}=4$ SYM $SU(N)$ theory in four dimensions
\footnote{Another approach to study the highly excited string states in
$AdS_5\times S^5$ and their relation to specific sectors of dual gauge
theory has also been proposed in \cite{GKP}. For further study in this
direction see \cite{GKPall}.}. Such an
identification is very difficult for $AdS_5\times S^5$
itself mainly because the string theory in this maximally
supersymmetric background has not proved to be exactly solvable
yet.

Regarding the fact that the string theory on a PP-wave background,
obtained by taking the Penrose limit of $AdS_5\times S^5$, is
exactly solvable one might wonder if this is the case for other
gravity backgrounds of the string theory. In particular if this also
works for string theory on PP-wave coming from the Penrose limit
of a gravity backgrounds with less than maximal supersymmetries.
In fact it has been observed that taking the Penrose limit of a
type IIB gravity background with the form of $AdS_5\times {\cal M}_5$,
where ${\cal M}_5$ is a five dimensional Einstein manifold which
could also be singular and therefore with less supersymmetries,
would lead to a PP-wave background in which the string theory can
be exactly solved. We note however that the PP-wave background could
be either the maximal supersymmetric
PP-wave, the same as one obtained by
$AdS_5\times S^5$, or a PP-wave background with less than
32 supercharges. For more detail of these observations see
\cite{all}.

Of course, it is not always the case that a PP-wave limit of a
gravity solution leads to a background in which the string theory
can be exactly solved (for example see \cite{HRV})\footnote{The
Penrose limit of different gravity backgrounds have also been studied
in \cite{all2}.}. Nevertheless there are examples of PP-wave systems
which provide a class of models that are exactly solvable in
string perturbation theory \cite{{RT},{Warner},{Myers}}.
In \cite{holog}, the holography problem, and in
\cite{{DP},{all3}}
open strings and D-branes have also been studied in the PP-wave background.

The PP-wave solutions we have been talking so far are mostly type IIB
PP-wave backgrounds. A way to find type IIA PP-wave is to use T-duality.
In fact in \cite{MIC} a supersymmetric type IIA PP-wave background has
found by making use of T-duality on the maximally supersymmetric type IIB
PP-wave. It has also been shown that the obtained type IIA PP-wave
preserves only 24 supercharges. The PP-waves with 24 and 28 supercharges in
IIB have also been studied in \cite{BR}. More recently a unique type IIA
PP-wave with 26 supercharges has been presented in \cite{MIC2} where the
authors argued that there is no type IIA PP-waves preserving 28 or 30
supercharges. This argument is based on the fact that there is no
11-dimensional PP-wave with 28 and 30 supercharges and there is a unique
11-dimensional PP-wave with 26 supercharges. Therefore upon compactification,
we just get PP-wave with 26 supercharges. Moreover it was argued that there is
no supersymmetric D-branes in this type IIA background.

This is the aim of this paper to study type IIA string theory on the PP-wave
background with 24 supercharges. In this case, we find that the theory can be
exactly solved and then quantized. We shall also see that the theory has
supersymmetric Dp-branes for $p=2,4,6,8$.

The organization of the paper is as following. In section 2 we will
review the compactification and T-duality in type IIB string theory on
the maximally supersymmetric PP-wave background which leads to a type IIA
PP-wave background with 24 supercharges. In section 3 we shall study the
bosonic and fermionic closed string in this background. In section 4 we
work out the first quantized bosonic and fermionic Hamiltonian of the system.
In section 5 the bosonic and fermionic open string sectors in this background
are studied. Finally the last section is devoted to the conclusions
and comments.

\section{PP-wave solution in type IIA}

The supergravity solution of the maximally supersymmetric PP-wave
in type IIB string theory is given by \cite{BFHP}
\bea
dS^2&=&2dX^+dX^--\mu^2X^iX^i (dX^+)^2+dX^idX^i,\cr
&&F_{+1234}=F_{+5678}={1\over 2}\mu,
\label{PPB}
\eea
where
$i=1,2,\cdots, 8$. This solution has $SO(4)\times SO(4)\times Z_2$ symmetry
which rotates and exchanges the $\{X^1,X^2,X^3,X^4\}$ and
$\{X^5,X^6,X^7,X^8\}$ subspaces. Its 30 Killing vectors are given by
\cite{BFHP}
\be\ba {ll}
k_{e_i}=-\cos(2\mu x^+)\partial_i-2\mu\sin(2\mu x^+)x^i\partial_-\;,
&k_{e_+}=-\partial_+,\cr &\cr
k_{e^*_i}=-2\mu\sin(2\mu x^+)\partial_i+
4\mu^2\cos(2\mu x^+)x^i\partial_-\;, &k_{e_-}=-\partial_-,\cr &\cr
k_{M_{ij}}=x^i\partial_j
-x^j\partial_i,\;\;\;\;\;\;\;\;\;\;\;{\rm both}\;\;i,j=1\cdots 4 & {\rm or}\;\;
 i,j=5\cdots 8,
\ea\ee

The compactification of the PP-wave background (\ref{PPB}) along the
circle generated
by the isometry $k_{e_-}+k_{e_+}$ has been considered in \cite{BFHP}.
Although this is
a manifest isometry of (\ref{PPB}), it is not space-like and moreover breaks all
supersymmetries, except at special values of radius of compactification.
On the other hand, this background has several isometries which are not manifest
in the form written in (\ref{PPB}). For example in \cite{MIC} the
compactification
of the background (\ref{PPB}) on the other directions has been considered. The
space-like of unit norm isometry, considered in \cite{MIC}, is given by
\be
k_{S^{\pm}_{ij}}=k_{e_i}\pm {1\over 2\mu}\;k_{e^*_j}\;.
\label{ISOM}
\ee
Using $SO(4)\times SO(4)\times Z_2$ symmetry of background one can choose the
``+'' sign and $i=1$. Then we left with two possibilities for $j$; either
$j=2$ or $j=5$. It was also shown that using $k_{S^+_{12}}$ the
compactified theory
would preserve only 24 supercharges. This is the case we work with in
this paper.

To make this isometry manifest, it is useful to change the
coordinates as following
\bea
X^+&=&x^+,\;\;\;\;\;X^-=x^--\mu
x^1x^2,\;\;\;\;\;X^I=x^I,\;\;{\rm for}\;I=3\cdots 8,\cr &&\cr
X^1&=&x^1\cos(\mu x^+)-x^2\sin(\mu x^+),\;\;\;\;X^2=x^1\sin(\mu
x^+)+x^2\cos(\mu x^+).
\eea
Then the solution (\ref{PPB}) reads
\bea
ds^2&=&2dx^+dx^--\mu^2 x^Ix^I (dx^+)^2-4\mu
x^2dx^1dx^++dx^idx^i\;,\cr &&\cr
F&=&{\mu\over 2}dx^{+}(dx^1dx^2dx^3dx^4+dx^5dx^6dx^7dx^8)\;.
\eea
In this coordinate system, ${\partial \over\partial x_1}$ is a
manifest isometry. Therefore we can compactify the background in
$x^1$ direction and perform T-duality along it. Doing so, we will
get the following type IIA PP-wave configuration \cite{MIC}
\bea
ds^2&=&2dx^+dx^--\mu^2[4(x^2)^2+(x^I)^2]dx^+dx^++dx^idx^i,\cr
B_{+1}&=&-2\mu x^2,\;\;\;\;\;F_{+234}=4\mu.
\label{PPA}
\eea
One can check that this solution solves the type IIA supergravity
equations of motion. To see this we note that the $\mu$ dependence
appears nontrivially only in the Ricci tensor $R_{++}$, where one
obtains a constant contribution of $10\mu^2$, which is canceled by
$2\mu^2$ contribution coming from B field and $8\mu^2$
contribution from RR 4-form.

This type IIA PP-wave background has $SO(2)\times SO(4)$ symmetry and preserves
24 supercharges. Its Killing spinor equation is given by
\be
(\nabla_{\mu}+ \Omega_{\mu}) \epsilon=0\;,
\ee
where
\be
\Omega_{\mu} = -{1\over 8} \Gamma^{\nu\rho} H_{\mu\nu\rho} \sigma_3 +
{1\over 8}{1\over 4!} F_{\nu\rho\lambda\delta}\Gamma^{\nu\rho\lambda\delta}
\Gamma_{\mu} \sigma_1\;.
\ee
For the solution (\ref{PPA}) one finds
\be\ba {ll}
\Omega_+ = {1\over 2}\mu \Gamma^{12} \sigma_3 +
\mu \Gamma^{+234} \Gamma_+ \sigma_1\;,  &\;\;\;\Omega_a = \mu \Gamma^{+234}
\Gamma_a \sigma_1\;,   \;\;\;\;\; a=3\cdots 8\;,\cr &\cr
\Omega_1 = -{1\over 2}\mu \Gamma^{+2} \sigma_3 +
\mu \Gamma^{+234} \Gamma_1 \sigma_1\;,  &\;\;\;
\Omega_2 = {1\over 2}\mu \Gamma^{+1} \sigma_3 +
\mu \Gamma^{+234} \Gamma_2 \sigma_1 \;,
\ea\ee
and the Killing spinor is given by
\be
\epsilon(\psi)=(1-\sum_{j=1}^{8} x^j \Omega_j)\chi(x^{+}),
\label{KILL}
\ee
where $\chi(x^+)={\rm e}^{-\Omega_+ x^+} \psi$ with
$\psi$ is a constant spinor. Regarding the fact
that $\chi$ depends only on $x^+$ gives the following two
equations,
\bea
&\partial_+ \chi + \Omega_+ \chi =0&\cr
&\sum_{j=1}^{8} \left(\delta_{ij} [ \Omega_j, \Omega_+] +
{1 \over 2} \mu^2 A_{ij}
\Gamma^- \Gamma^j \right) \chi = 0&\;,
\label{condition}
\eea
where $A_{ij}={\rm diag}(0, 4, 1, 1, 1, 1, 1, 1)$.
We note also that the solution (\ref{KILL}) should be
accompanied by the condition (\ref{condition}), which implies that
$\Gamma^+ \chi=\Gamma^+ \psi= 0$. Finally as it has been argued in \cite{MIC},
using this information one can show that the solution (\ref{PPA})
preserves only 24 supercharges.


\section{Type IIA string theory on PP-wave background}

In this section we shall study type IIA PP-wave on the PP-wave background
(\ref{PPA}). In comparison with type IIB PP-wave, this background has 6
massive bosonic fields plus a non-zero B field in two other directions.
Although in this case the equations of motion coming from
the Green-Schwarz action in light-cone gauge are coupled, we will see that
in fact they can exactly be solved and quantized. In this section, we shall
only consider the classical solution and leave the quantization to the
next section. To make the results more clear we will study the bosonic and
fermionic sectors separately.

\subsection{Bosonic sector}

The bosonic 2-dimensional worldsheet action in the presence of non-zero
B-field is given by
\be
S_{\rm bos}=-{1\over 4\pi\alpha'}\int
d\sigma^2\left[\eta^{ab}G_{\mu\nu}
\partial_a x^{\mu}\partial_b x^{\nu}+\epsilon^{ab}B_{\mu\nu}
\partial_a x^{\mu}\partial_b x^{\nu}\right],
\ee
where $\eta={\rm diag}(-1,1)$ is the worldsheet metric and we use a
notation in which $\epsilon^{\tau\sigma}=1$.

Plugging the type IIA PP-wave solution (\ref{PPA}) in this action, one
finds
\bea
S_{\rm bos}&=&-{1\over 4\pi \alpha'}\int
d\sigma^2\bigg{[}\eta^{ab}\left( 2\partial_a x^+\partial_b
x^-+\partial_a x^{i}\partial_b x^{i}-
\mu^2[4(x^2)^2+(x^I)^2]\partial_a x^{+}\partial_b x^{+}\right)\cr
&&\cr &-&2\mu x^2\epsilon^{ab}\left(\partial_a x^{+}\partial_b
x^{1}-\partial_a x^{1}\partial_b x^{+}\right)\bigg{]}\;.
\label{PPAA}
\eea
The same as type IIB case \cite{MET}, the action can be simplified
using the light-cone gauge. In this gauge, setting,
$x^+=p^+\tau$, the action (\ref{PPAA}) reads
\be
S_{\rm bos}=-{1\over 4\pi \alpha'} \int
d\sigma^2\left(\eta^{ab}\partial_a x^i\partial_b x^i+
m^2[(x^I)^2+4(x^2)^2]-4m x^2\partial_{\sigma} x^1\right),\label{abos}
\ee
where $m=\mu p^+$. From the variation of this action, the equations of motion
therefore in the light-cone gauge read
\bea
&&\eta^{ab}\partial_a\partial_b
x^I-m^2 x^I=0,\;\;\;\;\;{\rm for}\;\;I=3\cdots 8,\cr &&\cr &&
\eta^{ab}\partial_a\partial_b x^2-4m^2
x^2+2m\partial_{\sigma}x^1=0,\cr &&\cr
&&\eta^{ab}\partial_a\partial_b x^1-2m \partial_{\sigma}x^2=0
\eea
subject to the following boundary conditions
\be
\partial_{\sigma}x^I\delta x^I|_{\rm boundary}=0,\;\;\;
\partial_{\sigma}x^2\delta x^2|_{\rm boundary}=0,\;\;\;
\partial_{\sigma}x^1\delta x^1-2m x^2\delta x^1|_{\rm boundary}=0\,.
\label{bond}
\ee
Now we can proceed to solve the equation of motion subject to the
above boundary condition. In this section we only study the closed
string and leave the open string for the next section. The closed string
boundary conditions are given by
\be
x^I(\tau,\sigma+2\pi)=x^I(\tau,\sigma),\;\;\;
x^1(\tau,\sigma+2\pi)=x^1(\tau,\sigma),\;\;\;
x^2(\tau,\sigma+2\pi)=x^2(\tau,\sigma).
\ee
Using these boundary conditions one can consider an ansatz for the solution
as $e^{i(\omega \tau+n\sigma)}$ and the solution is then given by the following
mode expansion
\bea
x^1&=&x_0^1+\alpha'p_0^1\tau
+i\sqrt{\alpha'\over2}\sum_{n\neq0}{1\over \omega_n^1
}\left(\alpha_n^1e^{in\sigma}+\widetilde{\alpha}_n^1e^{-in\sigma}\right)
e^{-i\omega_n^1\tau}+\cr
&+&i\sqrt{\alpha'\over2}\sum_{n\neq0}{1\over \omega_n^2
}\left(\alpha_n^2e^{in\sigma}+\widetilde{\alpha}_n^2e^{-in\sigma}\right)
e^{-i\omega_n^2\tau}\;,\cr
x^2&=&x_0^2\cos(2m\tau)+{\alpha'\over 2m}p_0^2\sin(2m\tau)
+\sqrt{\alpha'\over2}\sum_{n\neq0}{{\rm sgn}(n)\over
n}\left(\alpha_n^1e^{in\sigma}-\widetilde{\alpha}_n^1e^{-in\sigma}\right)
e^{-i\omega_n^1\tau}+\cr &
+&\sqrt{\alpha'\over2}\sum_{n\neq0}{{\rm sgn}(n)\over
n}\left(\alpha_n^2e^{in\sigma}-\widetilde{\alpha}_n^2e^{-in\sigma}\right)
e^{-i\omega_n^2\tau}\cr
x^I&=&x_0^I\cos(m\tau)+{\alpha'\over
m}p_0^I\sin (m\tau)+i\sqrt{\alpha'\over2}\sum_{n \neq
0}{1\over \omega_n}(\alpha_n^I e^{in\sigma}+\widetilde{\alpha}_n^I
e^{-in\sigma})e^{-i\omega_n\tau} \;,
\eea
where
\be\ba {llll}
\omega_n^1=m+\sqrt{m^2+n^2}\;, &
\omega_n^2=m-\sqrt{m^2+n^2}\;, &
\omega_n=\sqrt{m^2+n^2}\,, & n>0 \cr & & &\cr
\omega_n^1=-m-\sqrt{m^2+n^2}\;, &
\omega_n^2=-m+\sqrt{m^2+n^2}\;, & \omega_n=-\sqrt{m^2+n^2}\;,
 & n<0.
\ea\ee
One should also impose the reality condition of fields which leads
to the following relations between the coefficients of Fourier modes
\be\ba {lll}
{\alpha}^{1\dag}_n=\alpha^1_{-n}\;,&
{\alpha}^{2\dag}_n=\alpha^2_{-n}\;, &
{\alpha}^{I\dag}_n=\alpha^I_{-n}\;,\cr &&\cr
\widetilde{\alpha}^{1\dag}_n=\alpha^1_{-n}\;, &
\widetilde{\alpha}^{2\dag}_n=\alpha^2_{-n}\;, &
\widetilde{\alpha}^{I\dag}_n=\alpha^I_{-n}\;.
\ea\ee

\subsection{Fermionic sector}

Probably the most interesting and also confusing part of the
Green-Schwarz action is the quadratic fermionic
part. Actually there are several expressions for the
quadratic fermionic part of the Green-Schwarz
action in a general background for type IIB string theory but
with several inconsistencies in signs and numerical factors
(for example see \cite{Warner}).

An expression for the quadratic fermionic part of the Green-Schwarz
action in a general background in type IIA is also given in
\cite{CLPS}. It seems to us that this action has the same problem
as one in type IIB, namely the correct coefficients seem to be different
from that appeared for example in equation (2.10) in \cite{CLPS}.
Indeed, following \cite{Warner} one could find a clear principle
that defines the relevant part of the action \cite{MT}:
the differential operator is precisely the
supercovariant derivative that appears in the gravitino variation.
Using this procedure, the quadratic fermionic action in type IIA
is given by
\be
{i\over\pi}\int d^2\sigma\left(\eta^{ab}\delta_{pq}
-\epsilon^{ab}(\sigma_3)_{pq}\right)\partial_a
x^{\mu}{\bar \theta}^p\Gamma_{\mu}{\cal D}_b\theta^q, \label{FAC}
\label{ACF11}
\ee
where $\theta^p, \; p=1,2$ are two 10 dimensional spinors with
different chiralities, and  $\sigma_3={\rm diag}(1,-1)$. The
generalized covariant derivative, ${\cal D}$ is given by
\cite{{ROM},{Hassan}}
\be
{\cal D}_a=\partial_a+{1\over 4}\partial_a x^{\rho}
\left[(\omega_{\mu\nu\rho}- {1\over 2} H_{\mu\nu\rho}\sigma_3)
\Gamma^{\mu\nu}+{1\over 2\times
 4!} F_{\mu\nu\lambda\delta}
\Gamma^{\mu\nu\lambda\delta}\sigma_1\Gamma_{\rho}\right],
\label{ACF22}
\ee
moreover we have also added a factor of half in the last term. As we will
see this is needed to get the expected supersymmetric solution. We will
give a comment about this later.

In the light-cone gauge we set $x^+=p^+\tau,\;\Gamma^+\theta^p=0$,
 then the non-zero contribution to the (\ref{FAC}) comes only from
the term where both the ``external'' and ``internal'' $\partial_ax^{\mu}$
factors become $p^+\delta_+^{\mu}\delta_a^{0}$. Therefore the action in
the light-cone gauge reads
\be
-{ip^+\over\pi}\int
d\sigma^2\;{\bar\theta}^p\Gamma_+\bigg{(}\delta_{pq} {\cal
D}_{\tau}+(\sigma_3)_{pq} {\cal D}_{\sigma}\bigg{)}\theta^q,
\label{ACF1}
\ee
where the supercovariant derivatives are given by
\bea
{\cal D}_{\tau}&=&\partial_{\tau}+{1\over 4}p^+ \left[(\omega_{\mu\nu
+}- {1\over 2} H_{\mu\nu +}\sigma_3) \Gamma^{\mu\nu}+{1\over
2\times 4!} F_{\mu\nu\lambda\delta}
\Gamma^{\mu\nu\lambda\delta}\sigma_1 \Gamma_{+}\right],\cr &&\cr
{\cal D}_{\sigma}&=&\partial_{\sigma}\;
\label{SUCO}
\eea
here $\sigma_i$'s are Pauli matrices. Plugging (\ref{SUCO}) into
(\ref{ACF1}) and taking into account that $\Gamma^+\theta^p=0$ one
finds the following expression for the quadratic fermoinic action in the
light-cone gauge
\bea
-{ip^+\over\pi}\int &
d\sigma^2&\bigg{[}{\bar\theta}^1\Gamma^-\partial_{+}\theta^1
+{\bar\theta}^2\Gamma^-\partial_{-}\theta^2+{m\over 2}\left(
{\bar\theta}^1\Gamma^-\Gamma^{12}\theta^1-
{\bar\theta}^2\Gamma^-\Gamma^{12}\theta^2\right)\cr &&\cr &+&
m\left({\bar\theta}^1\Gamma^-\Gamma^{234}\theta^2+
{\bar\theta}^2\Gamma^-\Gamma^{234}\theta^1\right)\bigg{]}\;.
\eea
The equations of motion read
\be
\left(\partial_{+}+{m\over 2}\Gamma^{12}\right)\theta^1+ m I
\theta^2=0,\;\;\;\;\left(\partial_{-}-{m\over
2}\Gamma^{12}\right)\theta^2+ m I \theta^1=0\;,
\ee
where
$I=\Gamma^{234}$. The fermionic boundary condition is also
given by
\be
\partial_\sigma(\overline{\theta}^1\Gamma^-\delta\theta^1
+\overline{\theta}^2\Gamma^-\delta\theta^2)(\tau,\sigma=2\pi)=
\partial_\sigma(\overline{\theta}^1\Gamma^-\delta\theta^1
+\overline{\theta}^2\Gamma^-\delta\theta^2)(\tau,\sigma=0)
\label{bondfer}
\ee
which is satisfied by imposing the periodicity condition on the
fermions for closed string theory. The open string sector will
be considered in the next section.

Using an ansatz for the fermions as
$\theta e^{i(\omega\tau+n\sigma)}$
and ${\tilde \theta} e^{i(\omega\tau+n\sigma)}$ and imposing the
closed string boundary conditions, we get the following mode expansion for
the fermionic fields
\bea
\theta^1&=&e^{-{m\over2}\Gamma^{12}\tau}\left( \cos(m\tau)\theta_0+
\sin(m\tau)\widetilde{\theta}_0\right)\cr &&\cr
&+&\sum_{n\neq 0}\left(e^{-i(\omega_n\tau-n\sigma)}\theta_n+{im\over 2n}
e^{-i(\omega_n\tau+n\sigma)}\widetilde{\theta}_n\right)\;,\cr &&\cr
\theta^2&=&Ie^{-{m\over2}\Gamma^{12}\tau}\left(-\sin(m\tau)\theta_0 +
\cos(m\tau)\widetilde{\theta}_0\right)\cr &&\cr
&+&({-iI\over m})\sum_{n\neq 0}\bigg{[}\left(\sqrt{m^2+n^2}{\rm sgn}(n)-n
\right) e^{-i(\omega_n\tau-n\sigma)}\theta_n \cr &&\cr
&+&{im\over2n}\left(\sqrt{m^2+n^2}{\rm sgn}(n)+n\right)
e^{-i(\omega_n\tau+n\sigma)}\widetilde{\theta}_n\bigg{]}\;,
\eea
with
\bea
&&\omega_n=-{im\Gamma^{12}\over2}+\sqrt{m^2+n^2},\,\,\, n>0\;,\cr &&\cr
&&\omega_n=-{im\Gamma^{12}\over2}-\sqrt{m^2+n^2},\,\,\, n<0\;.
\eea
The reality condition for the fermionic field implies that
\be
{\theta}_n^{\alpha\dag}=\theta^{\alpha}_{-n},\,\,\,\,\,\
\widetilde{\theta}_n^{\alpha\dag}=\widetilde{\theta}^{\alpha}_{-n}
\ee
where $\alpha$ is the spinor index.

Since $i\Gamma^{12}$ has eigenvalues $\pm 1$, the fermionic frequencies
are the same as bosonic's up to an $m$ dependent constant. In fact
this is what we would expect to get as a necessary condition for the
supersymmetry. The situation is very similar to the
type IIB PP-wave compactified on a circle using the isometry $k_{S^+_{12}}$
in (\ref{ISOM}) where the system has 24 charges. This model has been
studied in \cite{MIC} where the relation between bosonic and fermionic
modes are the same as our case.

Note that taking another form for the quadratic fermionic part of the action, for
example by dropping the factor of half in the last term of supercovariant
derivative or that considered in \cite{CLPS} we would get another answer which
might not have a reasonable explanation for the supersymmetry which is presence
in the system. Regarding this fact, we believe the fermionic action we
started with,
(\ref{ACF11}) and (\ref{ACF22}) is the correct one.

\section{Quantization}

In this section we proceed to study the canonical quantization of our
model considered in the previous section. Again to make our computation
more clear we shall consider the bosonic and fermionic sectors
separately.

\subsection{Bosonic sector}

Using the bosonic mode expansion one can write an expression for the
canonical momentum ${\cal P}^i={\dot x}^i$ and then using the canonical
quantization relation
\be
[{\cal P}^i(\tau,\sigma), x^{j}(\tau,\sigma')]=\delta^{ij}
\delta(\sigma-\sigma'),
\ee
we get the following commutation relations
\be\ba {ll}
[a_r^I,a_s^J]=[\widetilde{a}_r^I,\widetilde{a}_s^J]
={\rm sgn}(r)\delta_{r+s,0}\;\delta^{IJ}\;, &
[x_0^1,p_0^1]=[x_0^2,p_0^2]=i\;,\cr &\cr
[a_r^1,a_s^1]=[\widetilde{a}_r^1,\widetilde{a}_s^1]
=-{\rm sgn}(r)\delta_{r+s,0}\;, &[x_0^I,p_0^J]=i\delta^{IJ}\;,
\cr &\cr
[a_r^2,a_s^2]=[\widetilde{a}_r^2,\widetilde{a}_s^2]
={\rm sgn}(r)\delta_{r+s,0}\;, &
[\overline{a}_0^2,a_0^2]=1\;,
\,\,\,\,[\overline{a}_0^I,a_0^I]=1\;,
\ea\ee
where $a_i^n$ and $\widetilde{a}_i^n$ are defined in terms of the
oscillators modes as following
\bea
&a_0^2={1\over2}(\alpha'p_0^2+2imx_0^2),
\;\;\;\;\;\;\;\;\;\;
\overline{a}_0^2={1\over2}(\alpha'p_0^2-2imx_0^2),&
\cr && \cr
& a_0^I={1\over\sqrt{2}}(\alpha'p_0^I+imx_0^I),
\;\;\;\;\;\;\;\;\;\;
\overline{a}_0^I={1\over\sqrt{2}}(\alpha'p_0^I-imx_0^I),&
\nonumber
\eea
\be\ba {lll}
a_r^2={\sqrt[4]{4(m^2+r^2)}\over ir}\alpha_r^2, &
a_r^1={\sqrt[4]{4(m^2+r^2)}\over ir}\alpha_r^1 , &
a_r^I={1\over\sqrt[4]{(m^2+r^2)}}\alpha_r^I ,
\cr & &\cr
\widetilde{a}_r^2={\sqrt[4]{4(m^2+r^2)}\over ir}\widetilde{\alpha}_r^2,
&\widetilde{a}_r^1={\sqrt[4]{4(m^2+r^2)}\over ir}\widetilde{\alpha}_r^1,&
\widetilde{a}_r^I={1\over\sqrt[4]{(m^2+r^2)}}\widetilde{\alpha}_r^I,
\ea\ee
The light-cone Hamiltonian of bosonic part is given by
\be
{\mathcal{H}}_b={1\over 4\pi\alpha'}\int_0^{2\pi}d\sigma
\left(({\cal P}^i)^2
+\partial_\sigma x^i\partial_\sigma x^i
+m^2[(x^I)^2+4(x^2)^2]-4mx^2\partial_\sigma x^1\right),
\ee
which in terms of creation and annihilation operators is
\bea
{\mathcal{H}}_b&=&4+{1\over2}(\alpha'p_0^1)^2
+2a_0^2\overline{a}_0^2+a_0^I\overline{a}_0^I
+\sum_{n\neq0}\omega_n{\rm sgn}(n)\left(a_{-n}^I a_n^I+
\widetilde{a}_{-n}^I\widetilde{a}_n^I\right)\cr &&\cr
&+&{1\over2}\sum_{n\neq0}\omega^1_n{\rm sgn}(n)\left(a_{-n}^1a_n^1+
\widetilde{a}_{-n}^1\widetilde{a}_n^1\right)
-{1\over2}\sum_{n\neq0}\omega^2_n{\rm sgn}(n)\left(a_{-n}^2a_n^2+
\widetilde{a}_{-n}^2\widetilde{a}_n^2\right) \label{hambos},
\eea
where $4$ in the Hamiltonian is the zero point energy, which can easily
be read from the zero-mode terms. In fact we have a contribution
${1\over2}\times 2$ from $x^2$ direction and ${1\over2}
\times 6$ form $x^I,\;I=3\cdots 8$ with altogether come to 4.

\subsection{Fermionic sector}

The canonical momenta of the fermionic fields are given by
\be
p^1_{\alpha}=ip^+ (\Gamma^-\theta^1)_{\alpha}\,\,\,\,\,,\,\,\,\,\,
p^2_{\alpha}=-ip^+ (\Gamma^-\theta^2)_{\alpha}\;.
\ee
The classical Poisson brackets between
fermionic fields and their conjugate momenta are
\be
\{p^i_{\beta},\theta^{j\alpha}\}=
{1\over2}{(\Gamma^{11}\Gamma^+\Gamma^-)
^{\alpha}}_{\beta}\delta^{ij}\delta(\sigma,\sigma')
\,\,\,\,\,,\,\,\,\,\,\Gamma^{11}=\Gamma^0\ldots\Gamma^9,
\ee
which lead to the classical Dirac brackets between fermionic fields
\be
\{{\theta}^{i\alpha},\theta^{j\beta}\}=
{1\over4ip^+}(\Gamma^+)^{\alpha\beta}\delta^{ij}\delta(\sigma,\sigma').
\ee
To quantize the theory we replace the classical brackets with
anticommutators for corresponding creation and annihilation
operators. Doing so, the anticommutation relations for fermionic
creation and annihilation operators become
\bea\ba {ll}
\{\vartheta^\alpha_r,\vartheta^\beta_s\}=
(\Gamma^{+})^{\alpha\beta}\delta _{r,-s}, &
\{\widetilde{\vartheta}^\alpha_r,\widetilde{\vartheta}^\beta_s\}=
(\Gamma^{+})^{\alpha\beta}\delta _{r,-s}\cr &\cr
\{\vartheta^\alpha_0,\vartheta^\beta_0\}=
(\Gamma^{+})^{\alpha\beta}, &
\{\widetilde{\vartheta}^\alpha_0,\widetilde{\vartheta}^\beta_0\}=
(\Gamma^{+})^{\alpha\beta}.
\ea\eea
where the normalized creation and annihilation operators
$\vartheta^{\alpha}_i$
and $\widetilde{\vartheta}^{\alpha}_i$ in terms of the mode
expansion coefficient
${\theta}^{\alpha}_i$ and $\widetilde{\theta}^\alpha_i$ are defined by
\bea\ba {ll}
\vartheta_r=\sqrt{{16p^+\sqrt{m^2+r^2}
\over \sqrt{m^2+r^2}+r{\rm sgn}(r)}}\theta_r , &
\widetilde{\vartheta}_r=\sqrt{{4p^+m^2\sqrt{m^2+r^2}
\over r^2(\sqrt{m^2+r^2}-r{\rm sgn}(r))}}\widetilde{\theta}_r\cr &\cr
\vartheta_0=\sqrt{8p^+}\theta_0 , &
\widetilde{\vartheta}_0=\sqrt{8p^+}\widetilde{\theta}_0\;.
\ea\eea

Using the equations of motion, the Hamiltonian for fermionic sector
takes the following  form
\be
{\mathcal{H}}_f=-{ip^+\over2\pi}
\int_0^{2\pi}d\sigma(\overline{\theta}^1\Gamma^{-}
\partial_{\tau}\theta^1+\overline{\theta}^2\Gamma^{-}
\partial_{\tau}\theta^2),
\ee
and in terms of creation and annihilation operators the Hamiltonian
reads
\bea
&&{\mathcal{H}}_f={im\over 16}\left(
\overline{\vartheta}_0\Gamma^{-}\Gamma^{12}\vartheta_0
+\overline{\widetilde{\vartheta}}_0
\Gamma^{-}\Gamma^{12}\widetilde{\vartheta}_0
-4\overline{\vartheta}_0\Gamma^{-}\widetilde{\vartheta}_0
\right)\cr &&\cr
&&-{1\over8}\sum_{n\neq0}\overline{\vartheta}_{-n}\Gamma^{-}
\omega_n\vartheta_n
-{1\over8}\sum_{n\neq0}\overline{\widetilde{\vartheta}}_{-n}\Gamma^{-}
\omega_n\widetilde{\vartheta}_n\;.
\label{hamfer}
\eea
It can also be seen that
the zero point energy is zero for the fermionic sector.

Note that the metric we started with has $SO(6)$ global symmetry while the
4-form RR field breaks the symmetry to $SO(2)\times SO(4)$. We might
expect to see this symmetry breaking in the fermionic zero mode. We note
however that whether the symmetry breaking is manifest in the
level of Hamiltonian depends on the way we define the creation and annihilation
operators. One could choose the the creation and annihilation
operators in such a way that the vacuum preserves $SO(6)$ symmetry, but
the zero mode of fermionic Hamiltonian is not $SO(6)$ invariant. On the
other hand, one could introduce another set of  creation and annihilation
operators which preserves only $SO(2)\times SO(4)$ global symmetry, but the
zero mode of the fermionic Hamiltonian will formally restores the full
$SO(6)$ symmetry. Here we have taken the second choice. For the similar
discussion for the type IIB on PP-wave background see \cite{MT}.

Finally we note also that, in the light-cone gauge where we set
$x^+=p^+\tau$, the $x^-$ coordinate can be fixed using the Virasoro
constraint. In fact we get the following constraint for $x^-$
\be
p^+\partial_{\sigma} x^-+\partial_{\sigma}x^i\partial_{\tau}x^i
-4i\alpha'p^+(\overline{\theta}^1\Gamma^-\partial_{\sigma}\theta^1
+\overline{\theta}^2\Gamma^-\partial_{\sigma}\theta^2)=0\;.
\ee
Plugging the mode expansions for bosonic and fermionic fields into this
constraint one finds the usual constraint for the left and right
moving modes appear in each level, namely
\be
{\mathcal{N}}_1={\mathcal{N}}_2,
\ee
where
\bea
{\mathcal{N}}_1&=&\pi\alpha'\sum_{n \neq0}\bigg[ n\,\,{\rm sgn}(n)\left(
-a^1_{-n}a^1_n+a^2_{-n}a^2_n-a^I_{-n}a^I_n\right)+n\overline{\vartheta}
_{-n}\Gamma^-\vartheta_n\bigg]\cr &&\cr
{\mathcal{N}}_2&=&\pi\alpha'\sum_{n \neq0}\bigg[ n\,\,{\rm sgn}(n)\left(
-\widetilde{a}^1_{-n}\widetilde{a}^1_n
+\widetilde{a}^2_{-n}\widetilde{a}^2_n
-\widetilde{a}^I_{-n}\widetilde{a}^I_n\right)
+n\overline{\widetilde{\vartheta}}_{-n}\Gamma^-\widetilde{\vartheta}_n\bigg],
\eea
are oscillation numbers for left and right moving sectors
in our closed string theory.

\section{Open strings}

So far, we have considered the closed string on  the PP-wave
background (\ref{PPA}). It is also important to study the open
string sector on this PP-wave background. This is because by making use
of the open string dynamics we can find and classify  D-branes of
the theory. The D-branes of type IIB string theory on the maximally
PP-wave background have been studied in \cite{{DP}, {all3}} where it
was shown that the theory has supersymmetric Dp-branes with half of the
original supersymmetries only for $p=3,5,7$. In this section we would like
to study D-branes of the type IIA string theory on the PP-wave
background (\ref{PPA}). Here we will use the notation of \cite{DP}.
We will separately consider the bosonic and fermionic sectors to make the
thing more clear.

\subsection{Bosonic sector}

In this subsection we study the bosonic open string solutions for
the action (\ref{abos}). In the study of open string the essential
role is played by the boundary conditions which we can impose
for different directions. This could be either Neumann or
Dirichlet boundary conditions. We note, however, since our background
(\ref{PPA}) has also a non-zero B field in
``1'' and ``+'' directions we could have mixed boundary condition
for these directions as well.

In the light-cone gauge where $x^+=p^+\tau$ the only consistent
boundary condition for $x^+$ is Neumann boundary condition. On the other hand
from the bosonic part of the Virasoro constraint we get
\be
\partial_{\sigma} x^-=-{\partial_{\sigma}x^i\partial_{\tau}x^i\over
p^+}
\ee
which means for $p^+\neq 0$ the only boundary condition for $x^-$ is
Neumann boundary condition too. Therefore the branes we are
considering in this section are along $x^+$ and $x^-$ directions.

For the direction $x^I$ we could choose either Neumann or Dirichlet boundary
conditions. For Neumann boundary condition in $x^I$ direction we get
\be
x^I=x_0^I\cos(m\tau)+{\alpha'p_0^I\over m}\sin(m\tau)+
i\sqrt{2\alpha'}\sum_{n\neq0}{\alpha_n^I\over \omega_n}
\cos(n\sigma)e^{-i\omega_n\tau}\;,
\ee
which gives the following contribution to the open string Hamiltonian
of bosonic sector
\be
{\mathcal{H}}_I=a_0^I{\overline{a}}_0^I+{N \over 2} +2\sum_{n\neq 0}\omega_{n}
{\rm sgn}(n)(a_{-n}^Ia_n^I)\;,
\ee
where $N/2$ is the zero point energy corresponding to $N$-Neumann directions.
On the other hand, for the Dirichlet boundary condition in $x^I$ direction
one finds
\be
x^I=-\sqrt{2\alpha'}\sum_{n\neq0}{\alpha_n^I\over \omega_n}
\sin(n\sigma)e^{-i\omega_n\tau}\;,
\ee
with the following contribution to the open string Hamiltonian in the
bosonic sector
\be
{\mathcal{H}}_I=2\sum_{n\neq0}\omega_{n}
{\rm sgn}(n)(a_{-n}^Ia_n^I)\;.
\ee
In the above equations, the frequency is defined the same as closed
string case, namely $\omega_n=\sqrt{m^2+n^2}\;{\rm sgn}(n)$.

For $x^2$ direction the situation is the same as $x^I$, though
with different mode expansion. On the other hand for $x^1$
direction we could have mixed boundary condition as following
\be
\partial_\sigma x^a+B_{b}^a\partial_\tau x^b=0,\;\;\;\;{\rm for}\;
a,b=+,1\;.
\ee
But since the boundary condition for $x^+$ is always Neumann, the
$x^1$ must obey the following boundary condition
\be
\partial_{\sigma}x^1-2m x^2|_{\rm boundary}=0\,.
\label{X1}
\ee
Looking at the mode expansion of $x^2$ we realize that for the
Dirichlet boundary condition the value of $x^2$ at the boundary must
be zero. From (\ref{X1}) this means that the boundary condition
for $x^1$ must be Neumann. On the other hand for the Neumann boundary
condition for $x^2$ one can show that the condition (\ref{X1}) is
not satisfied and therefore we need to set $\delta x^1=0$ at the boundary,
{\it i.e.} the boundary condition for $x^1$ must be Dirichlet.
Therefore choosing Neumann boundary condition for $x^2$ implies
that $x^1$ must have Dirichlet boundary condition and vise versa. In these
cases, their mode expansions are given by

{\bf 1}) {\it Neumann boundary condition for $x^2$}
\bea
x^2&=&x_0^2\cos(2m\tau)+{\alpha'p_0^2\over 2m}
\sin(2m\tau)+\cr &&\cr
&+&\sqrt{2\alpha'}\sum_{n\neq0}{{\rm sgn}(n)\over n}
\cos(n\sigma)\left(\alpha_n^1e^{-i\omega^1_n\tau}
+\alpha_n^2e^{-i\omega^2_n\tau}\right)\cr &&\cr
x^1&=&-\sqrt{2\alpha'}\sum_{n\neq0}\sin(n\sigma)
\left({\alpha_n^1\over \omega^1_n}
e^{-i\omega^1_n\tau}+{\alpha_n^2\over \omega^2_n}
e^{-i\omega^2_n\tau}\right)\;,
\eea
with the following contribution to the bosonic Hamiltonian
for open string
\be
{\mathcal{H}}_{1,2}=2a_0^2\overline{a}_0^2+1
+\sum_{n\neq0}\omega^1_n{\rm sgn}(n)\left(a_{-n}^1a_n^1\right)
-\sum_{n\neq0}\omega^2_n{\rm sgn}(n)\left(a_{-n}^2a_n^2\right)
\ee

{\bf 2}) {\it Dirichlet boundary condition for $x^2$}
\bea
x^2&=&i\sqrt{2\alpha'}\sum_{n\neq0}{{\rm sgn}(n)\over n}
\sin(n\sigma)\left(\alpha_n^1e^{-i\omega^1_n\tau}
+\alpha_n^2e^{-i\omega^2_n\tau}\right)\cr &&\cr
x^1&=&x_0^1+\alpha'p_0^1\tau+i\sqrt{2\alpha'}
\sum_{n\neq0}\cos(n\sigma)\left({\alpha_n^1\over \omega^1_n}
e^{-i\omega^1_n\tau}+{\alpha_n^2\over \omega^2_n}
e^{-i\omega^2_n\tau}\right)\;,
\eea
and its contribution to the bosonic Hamiltonian is
\be
{\mathcal{H}}_{1,2}={1\over2}(\alpha'p_0^1)^2+
\sum_{n\neq0}\omega^1_n{\rm sgn}(n)\left(a_{-n}^1a_n^1\right)
-\sum_{n\neq0}\omega^2_n{\rm sgn}(n)\left(a_{-n}^2a_n^2\right)\;.
\ee
In these equations the frequencies are defined the same as closed string
case
\be
\omega_n^1=(m+\sqrt{m^2+n^2})\;{\rm sgn}(n),\;\;\;\;\;\;
\omega_n^2=(m-\sqrt{m^2+n^2})\;{\rm sgn}(n)\;.
\ee

\subsection{Fermionic sector}
Since $\theta^1$ and $\theta^2$ have different chiralities, the
fermionic boundary condition (\ref{bondfer}) is satisfied for
\be
\theta^1=M\Gamma^0\theta^2,
\ee
where $M$ is made out of product of even number of ten dimensional
gamma matrices which satisfies the following conditions
\be
-M^T\Gamma^0\Gamma^-M\Gamma^0=\Gamma^-\;,\;\;\;\;\;
M\Gamma^0IM\Gamma^0I=-1.
\label{MDR}
\ee
Using this boundary condition for open fermionic string, we find
the following solution for the fermionic field
\bea
\theta^1&=&
e^{-{m\over2}\Gamma^{12}\tau}\left(\cos(m\tau)-
M\Gamma^0I\sin(m\tau)\right)\theta_0+\cr &&\cr
&&+\sum_{n\neq0}e^{-i\omega_n\tau}\left[\theta_n
e^{in\sigma}-{iM\Gamma^0I\over m}\left(\sqrt{m^2+n^2}{\rm sgn}(n)
-n\right)\theta_n e^{-in\sigma}\right]\;,\cr &&\cr
\theta^2&=&
-Ie^{-{m\over2}\Gamma^{12}\tau}\left(\sin(m\tau)+
M\Gamma^0I\cos(m\tau)\right)\theta_0+\cr &&\cr
&&+{iI\over m}\sum_{n\neq0}e^{-i\omega_n\tau}
\left[\left(\sqrt{m^2+n^2}{\rm sgn}(n)
-n\right)\theta_ne^{in\sigma}-{imM\Gamma^0I}
\theta_n e^{-in\sigma}\right]\;,
\eea
and the fermionic Hamiltonian for the fermionic sectors is given by
\bea
&&{\mathcal{H}}_f=
{im\over 16}\overline{\vartheta}_0\left(\Gamma^-
\Gamma^{12}-IM^T\Gamma^0\Gamma^-
\Gamma^{12}M\Gamma^0I+4\Gamma^-M\Gamma^0I\right)\vartheta_0-\cr &&\cr
&&-{1\over 8}\sum_{n\neq0}\overline{\vartheta}_{-n}
\left[\Gamma^-\omega_n-{4n^2\over m^4}
(\sqrt{m^2+n^2}{\rm sgn}(n)-n)^2IM^T\Gamma^0\Gamma^
-\omega_nM\Gamma^0I\right]\vartheta_n
\eea
where $\omega_n={-im\Gamma^{12}\over2}+\sqrt{m^2+n^2}\;
{\rm sgn}(n)$, the same as the closed string case.

To classify the possible supersymmetric D-branes in type IIA
string theory on the PP-wave background (\ref{PPA}), we should
solve the conditions (\ref{MDR}) for $M$. To do this, it is convenient
to define $\Omega$ as
$M=\gamma^0\Omega$, which is made out of product of odd
number of gamma matrices, {\it i.e.} $\Omega=\Gamma^{\mu_1}
\cdots\Gamma^{\mu_{2k+1}}$. These indices correspond to the
directions which have Dirichlet boundary condition, in other
words, the directions which are transverse to the brane. This
shows that we will have even branes.

It turns out that the only solutions for the conditions
(\ref{MDR}) are
\bea
\Omega&=&\Gamma^a\;,\cr
\Omega&=& \Gamma^i\Gamma^j\Gamma^k
,\,\,\,\,\,\,\,\,\,\,\,\,\,\,\,\,\,\,
\Omega= \Gamma^a\Gamma^b\Gamma^i\;,\cr
\Omega &=& \Gamma^2\Gamma^3\Gamma^4\Gamma^i\Gamma^j\;,
\,\,\,\,\,\,\,\,\Omega=
\Gamma^a\Gamma^i\Gamma^j\Gamma^k\Gamma^l\;,\cr
\Omega&=& \Gamma^a
\Gamma^b\Gamma^i\Gamma^j\Gamma^k\Gamma^l\Gamma^m\;,
\label{IIABR}
\eea
where $a,b,c=2,3,4$, and $i,j,k,l,m=1,5,6,7,8$. This means that we can have
supersymmetric Dp-branes with $p=2,4,6,8$. This is in fact what we would
expect from T-duality of corresponding D-branes in type IIB \cite{DP}.

\section{Conclusions}

In this paper we have studied type IIA string theory on a PP-wave
background which preserves 24 supercharges. This PP-wave background
can be obtained from the maximally supersymmetric PP-wave in type IIB
using T-duality \cite{MIC}. This model is exactly solvable. Of course
this is what might be expected, because the T-duality does not
change the quadratic structure of the action in the light-cone gauge,
though it could mix the fields to each others.

An essential fact about our consideration is how to write the type IIA
superstring action in a general background. Although the bosonic part has
a well-known form, there are several confusions for the quadratic
fermionic action. We have been able to fix this using the supercovariant
derivative which appears in the gravitino transformation in type IIA
superstring theory. This is analogues to that proposed in \cite{{MT},{
Warner}} for type IIB. We have found that using this procedure, the
solutions we get are consistent with the supersymmetry of theory.

In this paper we have also considered open string sectors and thereby the
possible supersymmetric D-branes in the theory. We found that type IIA in
the PP-wave background can not have D0-brane which of course can be
understood from the fact that both $x^+$ and $x^-$ have Neumann
boundary condition. Thus we could get Dp-brane for $p=2,4,6,8$. This is
of course consistent with the result of \cite{DP}. There the
authors have found that possible supersymmetric Dp-branes are those
with $p=3,5,7$. Starting from Dp-branes in type IIB \cite{DP} and
applying T-duality as in the section 2 we see that the possible type
IIA D-branes are those classified in (\ref{IIABR}).

\end{document}